\documentclass[MSNbibl,nameyear,dvips]{arxstspdf}
\usepackage{flushend}
\usepackage{stfloats}
\usepackage{graphicx}
\volume{29}
\issue{1}
\pubyear{2014}
\firstpage{113}
\lastpage{127}
\doi{10.1214/13-STS425} 

\makeatletter
\newtheorem{theorem}{Theorem}
\newtheorem{corollary}{Corollary}
\newtheorem{lemma}{Lemma}

\newproclaim{remark}{Remark}

\def\td{\stackrel{d}{\to}}
\def\tp{\stackrel{p}{\to}}
\def\done{\mathbf1_d}

\makeatother

\begin{document}
\begin{frontmatter}

\title{Test for a Mean Vector with Fixed or Divergent Dimension}%
\runtitle{ELM for Means}

\begin{aug}
\author[a]{\fnms{Liang} \snm{Peng}\corref{}\ead[label=e1]{peng@math.gatech.edu}},
\author[b]{\fnms{Yongcheng} \snm{Qi}\ead[label=e2]{yqi@d.umn.edu}}
\and
\author[c]{\fnms{Fang} \snm{ Wang}\ead[label=e3]{fang72\_wang@126.com}}
\runauthor{L. Peng, Y. Qi and F. Wang}

\affiliation{Georgia Institute of Technology,
University of Minnesota Duluth and Capital Normal University}

\address[a]{Liang Peng is Professor, School of Mathematics,
Georgia Institute of Technology,
Atlanta, Georgia 30332-0160, USA \printead{e1}.}
\address[b]{Yongcheng Qi is Professor,
Department of Mathematics and Statistics,
University of Minnesota Duluth,
1117 University Drive,
Duluth, Minnesota 55812, USA \printead{e2}.}
\address[c]{Fang Wang is Associate Professor,
School of Mathematical Sciences,
Capital Normal University,
Beijing 100048, PR China \printead{e3}.}

\end{aug}

%
\begin{abstract}
It has been a long history in testing whether a mean vector with a
fixed dimension has a specified value. Some well-known tests include
the Hotelling $T^2$-test and the empirical likelihood ratio test
proposed by Owen [\textit{Biometrika} \textbf{75} (1988) 237--249;
\textit{Ann. Statist.} \textbf{18} (1990) 90--120]. Recently, Hotelling
$T^2$-test has been modified to work for a high-dimensional mean, and
the empirical likelihood method for a mean has been shown to be valid
when the dimension of the mean vector goes to infinity. However, the
asymptotic distributions of these tests depend on whether the dimension
of the mean vector is fixed or goes to infinity. In this paper, we
propose to split the sample into two parts and then to apply the
empirical likelihood method to two equations instead of d equations,
where d is the dimension of the underlying random vector. The
asymptotic distribution of the new test is independent of the dimension
of the mean vector. A simulation study shows that the new test has a
very stable size with respect to the dimension of the mean vector, and
is much more powerful than the modified Hotelling $T^2$-test.
\end{abstract}

%
\begin{keyword}
\kwd{Empirical likelihood}
\kwd{high-dimensional mean}
\kwd{test}
\end{keyword}

\end{frontmatter}

\section{Introduction}
\label{intro} Suppose
$X_1=(X_{1,1},\ldots,X_{1,d})^T,\ldots,X_n=(X_{n,1},\allowbreak\ldots,X_{n,d})^T $
are independent random vectors having common distribution function $F$
with mean $\mu$ and covariance matrix $\Sigma$. It has been a long
history to test $H_0\dvtx  \mu=\mu_0$ against $H_a\dvtx  \mu\neq\mu_0$ for a
given $\mu_0$. When the dimension $d$ is fixed, a traditional test is
the so-called Hotelling $T^2$-test defined as
\begin{eqnarray*}
T^2&=&(\bar X_n-\mu_0)^T
\\
&&{}\cdot\Biggl\{
\frac1{n-1}\sum_{i=1}^n(X_i-
\bar X_n) (X_i-\bar X_n)^T \Biggr
\}^{-1}\\
&&{}\cdot(\bar X_n-\mu_0),
\end{eqnarray*}
where $\bar X_n=\frac1n\sum_{i=1}^nX_i$.
Another commonly used one is the empirical likelihood ratio test
proposed by Owen (\citeyear{Owe88}, \citeyear{Owe90}). More specifically, by defining the
empirical likelihood function as
%
\begin{eqnarray}
\label{old} L(\mu)&=&\sup\Biggl\{\prod_{i=1}^n
(np_i)\dvtx  p_1\ge0,\ldots,p_n\ge0,\nonumber\\[-8pt]\\[-8pt]
&&\hspace*{45pt}\sum
_{i=1}^np_i=1, \sum
_{i=1}^np_iX_i=\mu\Biggr\},\nonumber
\end{eqnarray}
Owen (\citeyear{Owe88}, \citeyear{Owe90}) showed that the Wilks theorem holds under some
regularity conditions, that is, $-2\log L(\mu_0)$ converges in
distribution to a chi-square limit with $d$ degrees of freedom,
where $\mu_0$ denotes the true value of the mean of $X_i$.
Therefore, based on the chi-square limit, one can construct a
confidence region for $\mu$ or test $H_0\dvtx \mu=\mu_0$ against
$H_a\dvtx \mu\neq\mu_0$.

Without assuming a family of
distributions for the data, the empirical likelihood ratio
statistics can be defined to share similar properties as the
likelihood ratio for parametric distributions. For instance, the
empirical likelihood method produces confidence regions whose shape
and orientation are determined\vadjust{\goodbreak} entirely by the data. In comparison
with the normal approximation method and the bootstrap method for
constructing confidence regions, the empirical likelihood method
does not require a pivotal quantity, and it has better small sample
performance (see \cite{HalLaS90}). For more details on
empirical likelihood methods, we refer to \citet{Owe01} and the recent
review paper of \citet{CheVan09}.

Motivated by applications in neuroimaging and bioinformatics studies,
some tests for a mean vector with divergent dimension have been
proposed in the literature. It is known that, as the dimension is
large, the calculation of the inverse of the sample covariance matrix
in Hotelling $T^2$-test statistic becomes problematic and the sample
covariance matrix may diverge when $d/n\to c>0$; see
\citet{YinBaiKri88}. Moreover, Hotelling $T^2$-test is valid only
when $d<n$. In order to allow $d>n$, one may remove the sample matrix
in Hotelling's $T^2$-test statistic and avoid the singularity of the
sample covariance. This is exactly what has been done in
\citet{BaiSar96} and \citet{CheQin10} for the two-sample test
problem. The one-sample analogues of the two-sample test statistics in
\citet{BaiSar96} and \citet{CheQin10} lead to the following
test statistics:
\begin{eqnarray*}
M_n&=&(\bar X_n-\mu_0)^T (\bar
X_n-\mu_0)\\
&&\hspace*{0pt}{}-n^{-1} \operatorname{tr}\Biggl(\frac
1{n-1}\sum_{i=1}^n(X_i-\bar
X_n) (X_i-\bar X_n)^T \Biggr)
\end{eqnarray*}
and
\[
F_n=n^{-1}(n-1)^{-1}\sum
_{i\neq j}^n(X_{i}-\mu_0)^T
(X_{j}-\mu_0),
\]
respectively, where $\operatorname{tr}$ means the trace of a matrix. It is
easy to check that
\begin{eqnarray*}
M_n&=&\frac1{n^2}\sum_{i=1}^n
\sum_{j=1}^n(X_i-
\mu_0)^T(X_j-\mu_0)
\\
&&{} -\frac1{n(n-1)}\\
&&\hspace*{12pt}{}\cdot\operatorname{tr}\Biggl\{\sum_{i=1}^n(X_i-
\mu_0) (X_i-\mu_0)^T\\
&&\hspace*{30.5pt}{}-n^{-1}
\sum_{i=1}^n\sum
_{j=1}^n(X_i-\mu_0)
(X_j-\mu_0)^T\Biggr\}
\\
&=&\frac1{n^2}\sum_{i=1}^n\sum
_{j=1}^n(X_i-
\mu_0)^T(X_j-\mu_0)
\\
&&{} -\frac1{n(n-1)}\\
&&\hspace*{12pt}{}\cdot\Biggl\{\sum_{i=1}^n(X_i-
\mu_0)^T(X_i-\mu_0)\\
&&\hspace*{23.5pt}{}-n^{-1}
\sum_{i=1}^n\sum
_{j=1}^n(X_i-\mu_0)^T(X_j-
\mu_0)\Biggr\}
\\
&=&\frac1{n(n-1)}\sum_{i=1}^n\sum
_{j=1}^n(X_i-\mu_0)^T(X_j-
\mu_0)\\
&&\hspace*{0pt}{}-\frac1{n(n-1)}\sum_{i=1}^n(X_i-
\mu_0)^T(X_i-\mu_0)
\\
&=&F_n.
\end{eqnarray*}
That is, the test in \citet{BaiSar96} is the same as that in
\citet{CheQin10} when the one-dimensional data is concerned. As
mentioned in the end of Section~3 of \citet{CheQin10}, the
asymptotic behavior of $F_n$ depends on whether $d$ is fixed or goes to
infinity. Alternatively, \citet{SriDu08} and \citet{Sri09}
proposed to replace the covariance matrix in Hotelling $T^2$-test
statistic by a diagonal matrix. Rates of convergence for the
high-dimensional mean are studied by \citet{KueVid10}. For
nonasymptotic studies, we refer to Arlot, Blanchard and Roquain
(\citeyear{ArlBlaRoq10N1,ArlBlaRoq10N2}).

Although it is known that the empirical likelihood method performs
worse when the dimension $d$ is large and the sample size $n$ is not
large enough,
\citet{HjoMcKVan09} and \citet{ChePenQin09}
showed that the empirical likelihood method for
a fixed-dimensional mean is still valid when $d=d(n)\to\infty$ as
$n\to\infty$. More specifically, they showed under some regularity
conditions that
$(2d)^{-1/2}\{-2\log L(\mu_0)-d\}$ converges in distribution to a
standard normal if $d\to\infty$ as $n\to\infty$.
That is, the limiting distribution differs when the dimension of the
mean vector is fixed or diverges.

Now, the question is whether there exists a way to test $H_0\dvtx
\mu=\mu _0$ against $H_a\dvtx  \mu\neq\mu_0$ without distinguishing the
dimension of $\mu$ is finite or goes to infinity. Motivated by the
tests in \citet{BaiSar96} and \citet{CheQin10}, we propose to
apply the empirical likelihood method to the equation $E\{(X_1-\mu_0)^T
(X_2-\mu_0)\}=0$ instead of $EX_1=\mu_0$ for testing $H_0\dvtx
\mu=\mu_0$ against $H_a\dvtx \mu\neq\mu_0$, where $X_1$ and $X_2$ are
independent and identically distributed random vectors with mean $\mu$.
Although the equation $E\{(X_1-\mu_0)^T (X_2-\mu_0)\}=0$ is equivalent
to $H_0\dvtx  \mu=\mu_0$, a test only based on the equation
$E\{(X_1-\mu_0)^T (X_2-\mu_0)\}=0$ has a poorer\vspace*{1pt} power than
a test based on $EX_1=\mu_0$. The reason is that
$E\{(X_1-\mu_0)^T(X_2-\mu_0)\}/d=\delta^2/n$ instead of the standard
order $1/\sqrt n$ when $EX_1=\mu_0+\delta{\mathbf1}_d/\sqrt n$, where
$\done=(1,\ldots, 1)^T $ is a $d$-dimensional vector. To overcome this
issue so as to improve the test power, we propose to add one more
linear equation. More specifically, we propose to consider the
following two equations:
\[
E\bigl\{(X_1-\mu_0)^T(X_2-
\mu_0)\bigr\}=0
\]
and
\[
E\bigl\{\done^T(X_1+X_2-2
\mu_0)\bigr\}=0.
\]
It is easy to see that $E\{\done^T(X_1+X_2-2\mu_0)\}/d=O(1/\sqrt n)$
rather than $O(1/n)$ when $EX_1=\mu_0+\delta{\mathbf1}_d/\sqrt n$. The
first equation ensures the consistency of the proposed test, and the
second equation enhances the power in detecting a deviation. It turns
out that the empirical likelihood method based on the above two
equations works for either fixed $d$ or divergent $d$. This differs
from the results in \citet{BaiSar96} and \citet{CheQin10}.
More interestingly, the new method allows one to easily include more
independent equations if such equations characterize the departure from
the null hypothesis and are available. On the other hand, when the
number of equations becomes large, the minimization in the empirical
likelihood method turns out to be nontrivial.

We organize this paper as follows. In Section~\ref{main}
the new methodology and main results are given. Section~\ref{sim}
presents a simulation study. All proofs are given in
Section~\ref{proof}.

\section{Methodology}\label{main}

Assume $X_1=(X_{1,1},\ldots,X_{1,d})^T,\ldots,
X_n=(X_{n,1},\allowbreak\ldots,X_{n,d})^T $ are independent and identically
distributed random vectors having common distribution function $F$ with
mean $\mu$ and covariance matrix $\Sigma$. For testing $H_0\dvtx \mu=\mu_0$
against $H_a\dvtx \mu\neq\mu_0$ for a given $\mu_0$, we propose to apply the
empirical likelihood method to the equations $E\{(X_1-\mu_0)^T
(X_2-\mu_0)\}=0$ and $E\{\done^T (X_1+X_2-2\mu_0)\}=0$. In order to
have two independent samples, we simply split the first $m=[n/2]$
observations into a subsample and the second $m$ observations into
another subsample, and put $Y_i(\mu)=(u_i(\mu), v_i(\mu))^T $, where
\begin{eqnarray*}
u_i(\mu)&=&(X_i-\mu)^T (X_{i+m}-\mu),\\
v_i(\mu)&=&\done^T (X_i+X_{i+m}-2
\mu) \quad\mbox{for } i=1,\ldots, m.
\end{eqnarray*}
Hence, $\{Y_i(\mu), 1\le i\le m\}$ are i.i.d. bivariate random
vectors. Define the empirical likelihood function as
%
\begin{eqnarray}
\label{new}\quad \tilde L(\mu)&=&\sup\Biggl\{\prod_{i=1}^m(mp_i)\dvtx
p_1\ge0,\ldots, p_m\ge0,\nonumber\\[-8pt]\\[-8pt]
&&\hspace*{39pt}\sum_{i=1}^mp_i=1, \sum_{i=1}^mp_iY_i(\mu)=0
\Biggr\}.\nonumber
\end{eqnarray}
By the Lagrange multiplier technique, we have $p_i=m^{-1}\{1+\beta^T
Y_i(\mu)\}^{-1}$ for $i=1,\ldots,m$ and
$\tilde l(\mu)=-2\log\tilde L(\mu)=2\sum_{i=1}^m\log\{1+\beta^T
Y_i(\mu)\}$, where $\beta=\beta(\mu)=(\beta_1(\mu), \beta_2(\mu
))^T $ satisfies
%
\begin{equation}
\label{Lag} \frac1m\sum_{i=1}^m
\frac{Y_i(\mu)}{1+\beta^T Y_i(\mu)}=0.
\end{equation}

Write $\Sigma=(\sigma_{i,j})_{1\le i\le d, 1\le j\le d}=E\{(X_1-\mu
)(X_1-\mu)^T \}$, the covariance matrix of $X_1$, and use
$\lambda_1\le\cdots\le\lambda_d$ to denote the $d$ eigenvalues of
the matrix $\Sigma$. Note that $\lambda_i$'s may depend on $n$ when
$d$ depends on $n$.

First we show the Wilks theorem under very general conditions.

\begin{theorem}\label{thm1}
Assume $\sum^d_{i=1}\sum^d_{j=1}\sigma_{i,j}>0$ and for some $\delta>0$
%
\begin{equation}
\label{momentu0}\quad \frac{E|u_1(\mu)|^{2+\delta}}{ (\sum^d_{i=1}\sum
^d_{j=1}\sigma_{i,j}^2 )^{(2+\delta)/2}}=o\bigl(n^{({\delta
+\min(\delta, 2)})/{4}}\bigr)
\end{equation}
and
%
\begin{equation}
\label{momentv0}\quad \frac{E|v_1(\mu)|^{2+\delta}}{ (\sum^d_{i=1}\sum
^d_{j=1}\sigma_{i,j} )^{(2+\delta)/2}}=o\bigl(n^{({\delta+\min
(\delta, 2)})/{4}}\bigr).
\end{equation}
Then under $H_0\dvtx  \mu=\mu_0$, $\tilde l(\mu_0)$ converges in
distribution to a chi-square limit with two degrees of freedom as $n\to
\infty$.
\end{theorem}

Based on the above theorem, one can test $H_0\dvtx \mu=\mu_0$ against
$H_a\dvtx  \mu\neq\mu_0$. A test with level $\alpha$ is to reject
$H_0$ when $ \tilde l(\mu_0)>\xi_{1-\alpha}$, where $\xi_{1-\alpha
}$ is the $(1-\alpha)$th quantile of a chi-square limit with two
degrees of freedom.

Note that the proposed method works as well if one is interested in
testing the difference of two mean vectors based on paired data.
However, it is not applicable to the two-sample case with different
sample sizes.

Next we verify Theorem \ref{thm1} by imposing conditions on the
moments and dimension of the random vector:

(A1): $0<C_1\le\liminf_{n\to\infty}\lambda_1\le
\limsup_{n\to\infty}\lambda_d\le C_2<\infty$ for some constants
$C_1$ and $C_2$;

(A2): For some $\delta>0$, $\frac1d\sum
^d_{i=1}E|X_{1,i}-\mu_i|^{2+\delta}=O(1)$; and

(A3): $d=o(n^{({\delta+\min(\delta,2)})/({2(2+\delta)})})$.

\begin{corollary}\label{cor1} Assume conditions \textup{(A1)}--\textup{(A3)}
hold. Then conditions (\ref{momentu0}) and (\ref{momentv0}) are
satisfied and,
thus, Theorem \ref{thm1} holds.
\end{corollary}

Condition (A3) is a somewhat restrictive condition for the dimension
$d$. Note that conditions (A1) and (A2) are related only to the
covariance matrix and higher moments on the components of the random
vectors. Condition (A3) can be removed for models with some special
dependence structures. For comparisons, we prove the Wilks theorem for
the proposed empirical likelihood method under the following model B
considered by \citet{BaiSar96}, \citet{ChePenQin09} and
\citet{CheQin10}:

\textit{Model} B. $X_i=\Gamma Z_i+\mu$ for $i=1,\ldots,n$, where
$\Gamma$ is a $d\times k$ matrix with $\Gamma\Gamma^T=\Sigma
=(\sigma_{i,j})$ and $Z_i=(Z_{i,1},\ldots,Z_{i,k})^T$ are i.i.d. random
$k$-vectors with $EZ_i=0$, $\operatorname{Var}(Z_i)=I_{k\times k}$,
$EZ_{i,j}^4=3+\Delta<\infty$ and $E\prod_{l=1}^kZ_{i,l}^{\nu_l}=
\prod_{l=1}^kEZ_{i,l}^{\nu_l} $ whenever $\nu_1+\cdots+\nu_k=4$
for nonnegative integers $\nu_l$'s.

\begin{theorem}\label{thm2} Assume\vspace*{1pt} $\sum^d_{i=1}\sum^d_{j=1}\sigma_{i,j}>0$.
Then under model B and $H_0\dvtx  \mu=\mu_0$, $\tilde l(\mu_0)$
converges in distribution to a chi-square limit with two degrees of
freedom as $n\to\infty$.
\end{theorem}

\begin{theorem}\label{thm3} Assume $\sum^d_{i=1}\sum^d_{j=1}\sigma
_{i,j}>0$ and put
\[
\tau=\frac{m\|\mu_0-\mu\|^4}{\sum_{i=1}^d\sum_{j=1}^d\sigma
_{i,j}^2}+\frac{2m({\mathbf1}_d^T(\mu_0-\mu))^2}{\sum_{i=1}^d\sum
_{j=1}^d\sigma_{i,j}}.
\]
Then under model B and $H_a\dvtx  \mu\neq\mu_0$, we have
%
\begin{equation}
\label{power} P\bigl(\tilde l(\mu_0)>\xi_{1-\alpha}\bigr)=P
\bigl(\chi^2_{2,\tau}>\xi_{1-\alpha}\bigr)+o(1)
\end{equation}
as $n\to\infty$, where
$\xi_{1-\alpha}$ denotes the $(1-\alpha)$th quantile of a chi-square
limit with two degrees of freedom, and $\chi^2_{2,\tau}$ denotes a
noncentral chi-square random variable with two degrees of freedom and
noncentrality parameter $\tau$.
\end{theorem}

\begin{remark}\label{rem1}
It can be seen from the proof of Theorem \ref{thm2} that assumption
$EZ_{i,j}^4=3+\Delta<\infty$ in model B can be replaced by the
much weaker condition $\max_{1\le j\le k}EZ_{1,j}^4=o(m)$.
\end{remark}

\begin{remark}\label{rem2}
Unlike \citet{BaiSar96} and \citet{CheQin10}, there is no
restriction on $d$ and $k$ for our proposed method in Theorem \ref
{thm2}. The only constraint imposed on matrix $\Gamma$ is also very
weak, that is, $\sum^d_{i=1}\sum^d_{j=1}\sigma_{i,j}>0$ or,
equivalently, $\sum^d_{i=1}X_{1,i}$ is a nondegenerate random variable.
\end{remark}

\begin{remark}
We notice that conditions (\ref{momentu0}) and (\ref{momentv0}) in
Theorem \ref{thm1} impose some restriction on $d$ implicitly. Whether
such a restriction can be relaxed or \mbox{removed} depends on how sharp the
moments in the right-hand sides of (\ref{momentu0}) and
(\ref{momentv0}) can be estimated. In Corollary~\ref{cor1}, since we
do not assume any dependence structure among the components of $X_1$,
the best order of $d$ allowed in (A3) is less than $n^{1/2}$
even for bounded~$X_1$. On the other hand, since model B
assumes that the components of $X_1$ are linear combinations of some
orthogonal random variables, conditions (\ref{momentu0}) and
(\ref{momentv0}) become trivial and, consequently, the restriction on
$d$ is removed in Theorem \ref{thm2}.
\end{remark}

\begin{remark}\label{rem4}
When the test in \citet{BaiSar96} is applied to model
B for one sample, its power is
%
\begin{equation}
\label{rem3} \Phi\biggl(-\xi^*_{1-\alpha}+\frac{n\|\mu_0-\mu\|^2}{\sqrt
{2\sum_{i,j=1}^d\sigma_{i,j}^2}}\biggr)+o(1),
\end{equation}
where $\Phi(x)$ denotes the standard normal distribution function and
$\xi_{1-\alpha}^*$ denotes its $(1-\alpha)$th quantile; see Theorem 4.1
of \citet{BaiSar96}. Under model B, assumption (A1), $d\to\infty$,
and $\mu=\mu_0+a_n\bar\mu$ for $a_n\neq0$ and $\|\bar\mu\|=1$, it
follows from Lemma~\ref{lem1} in Section~\ref{proof} that $\tau$ in
Theorem \ref{thm3} has the order $\Delta_1=\frac{na_n^4}d+\frac
{n(\done^T\bar\mu)^2a_n^2}d$, and the order of\vspace*{1pt}
$\frac{n\|\mu_0-\mu\|^2}{\sqrt{2\sum_{i,j=1}^d\sigma_{i,j}^2}}$ in
(\ref{rem3}) is $\Delta_2=\frac{na_n^2}{\sqrt d}$. When both
$\Delta_1\to\infty$ and $\Delta_2\to\infty$, the power of both tests
goes to one. Due to the o(1) term in Theorem \ref{thm3} and
(\ref{rem3}), one cannot claim which power goes to one faster in this
case. When $0<\lim\inf\Delta_2\le\lim\sup\Delta_2<\infty$ and $\frac
{(\done^T\bar\mu)^2}{\sqrt d}\to\infty$, the test in
\citet{BaiSar96} has a power bounded from one, but the proposed
new test has a power tending to one, that is, the proposed empirical
likelihood test is much more powerful than the test in
\citet{BaiSar96} in this situation. However, when
$0<\lim\inf\Delta_2\le\lim\sup\Delta _2<\infty$ and
$\frac{(\done^T\bar\mu)^2}{\sqrt d}\to0$, the proposed empirical
likelihood test is much less\vspace*{1pt} powerful. In this case, a different linear
functional $c^T(X_i+X_{i+m}-\mu_0)$ has to be employed to replace
$\done ^T(X_i+X_{i+m}-\mu_0)$ so as to improve the test power, where
$c$ is a $d$-dimensional constant. When $\done^T(X_i+X_{i+m}-\mu_0)$ is
replaced by any new\vspace*{1pt} functional $c^T(X_i+X_{i+m}-\mu_0)$ in the
definition of the empirical likelihood $\tilde L(\mu)$ given in
(\ref{new}), similar results to Theorems \ref{thm1}, \ref{thm2} and
\ref{thm3} can also be derived easily. Moreover, the above $\Delta_1$
becomes $\frac{na_n^4}d+\frac {n(c^T\bar\mu)^2a_n^2}d$. Therefore, when
$\frac{(\done^T\bar\mu )^2}{\sqrt d}\to0$, one can choose $c$ such that
$\lim\inf\frac {(c^T\bar\mu)^2}{\sqrt d}>0$ so as to improve the power.
However, as discussed in the \hyperref[intro]{Introduction}, it remains
open on how to find such $c$ or the optimal linear functionals.
\end{remark}

\section{Simulation Study}\label{sim}

We investigate the finite sample behavior of the proposed empirical
likelihood method (NELM) and compare
it with the Hotelling's $T^2$-test (HT)
and the test statistic $M_n$ in \citet{BaiSar96} (BS)
in terms of size and power. A simulation reveals that the standard
empirical likelihood method (OELM) in \citet{Owe90} has a size much
larger than the nominal level when $d>20$ and, thus, it makes no sense
to compare these two empirical likelihood methods.

Let $W_1,\ldots,W_d$ be independent and identically distributed
random variables with distribution function either the standard
normal [notation $N(0,1)$] or $t$ distribution with $6$ degrees of
freedom [notation $t(6)$]. Consider the following two models:
\begin{longlist}[\textit{Model} 2:]
\item[\textit{Model} 1:] $X_{1,1}=W_1+\delta/\sqrt n,
X_{1,2}=W_1+W_2+\delta/\sqrt n,\ldots, X_{1,d}=W_{d-1}+W_d+\delta
/\sqrt n$.
\item[\textit{Model} 2:]
\[
(X_{1,1},\ldots,X_{1,d})^T \sim N
\bigl(\delta{\mathbf1}_d/\sqrt n,\bigl(0.5^{|i-j|}\bigr)_{1\le i,j\le
d} \bigr),
\]
\end{longlist}
where $\delta\in R$ and $n$ is the sample size. The question is to test
$H_0\dvtx \mu=0$ against $H_a\dvtx  \mu\neq0$. Hence, the case of
$\delta =0$ denotes the size of tests. It is easy to check that these
two models are a special case of model~B in Section~\ref{main}. For
example, model 1 corresponds to model~B with
\[
\Gamma=\pmatrix{ &1&0&0&\cdots&0&0
\cr
&1&1&0&\cdots&0&0
\cr
&0&1&1&\cdots&0&0
\cr
&\cdot&\cdot&\cdot&\cdots&\cdot&\cdot
\cr
&0&0&0&\cdots&1&1}
\]
and\vspace*{1pt} $\sum_{i=1}^d\sum_{j=1}^d\sigma_{i,j}=4d-3$, $\sum_{i=1}^d\sum
_{j=1}^d\sigma_{i,j}^2=6d-5$. Hence, Theorem \ref{thm2} holds for model
1 without restriction on the dimension $d$, and $\tau$ in Theorem
\ref{thm3} equals
$\frac{md^2\delta^4}{(6d-5)n^2}+\frac{2md^2\delta^2}{(4d-3)n}$.
Moreover, $\Delta_1$ and $\Delta_2$ defined in Remark \ref{rem4} are
$d\delta ^2$ and $\sqrt d\delta^2$, respectively, as $d\to\infty$.
Hence, theoretically the proposed empirical likelihood method is much
more powerful when $\sqrt d\delta^2$ is bounded away from infinity.
When $\delta$ is fixed, both tests have a power tending to one. In this
case Theorem \ref{thm3} and equation (\ref{rem3}) in Remark \ref{rem4}
cannot be used to claim which test is more powerful theoretically, but
the simulation results below show that the empirical likelihood method
is still more powerful. Similarly, we can verify that Theorem
\ref{thm2} holds for model~2 without restriction on the\vspace*{1pt} dimension as
well and $\tau=\frac{md}{n^2}\frac
{\delta^4}{5/3-8(1-0.5^{2d})d^{-1}/9}+\frac{md}{n}\frac
{2\delta^2}{3-4(1-0.5^d)d^{-1}}$\vspace*{1pt} in Theorem~\ref{thm3}. Using Remark
\ref{rem4}, we conclude that the proposed empirical likelihood method
for model 2 is more powerful than the test in \citet{BaiSar96}
when $\sqrt d\delta^2$ is bounded away from infinity and $d\to \infty$.
When $\delta$ is fixed and $d\to\infty$, both tests have a power
tending to one and theoretical comparison does not exist. However, the
following simulation results show that the proposed empirical
likelihood method is more powerful.

By drawing $10\mbox{,}000$ random samples of sample size $n=100$ and $300$ from
$X=(X_{1,1},\ldots,X_{1,d})^T$ with $d=5,10,15,\ldots,200$ and
$\delta=0, 0.1, 0.5$, we calculate the empirical sizes
and powers of those tests mentioned above.

\begin{figure*}

\includegraphics{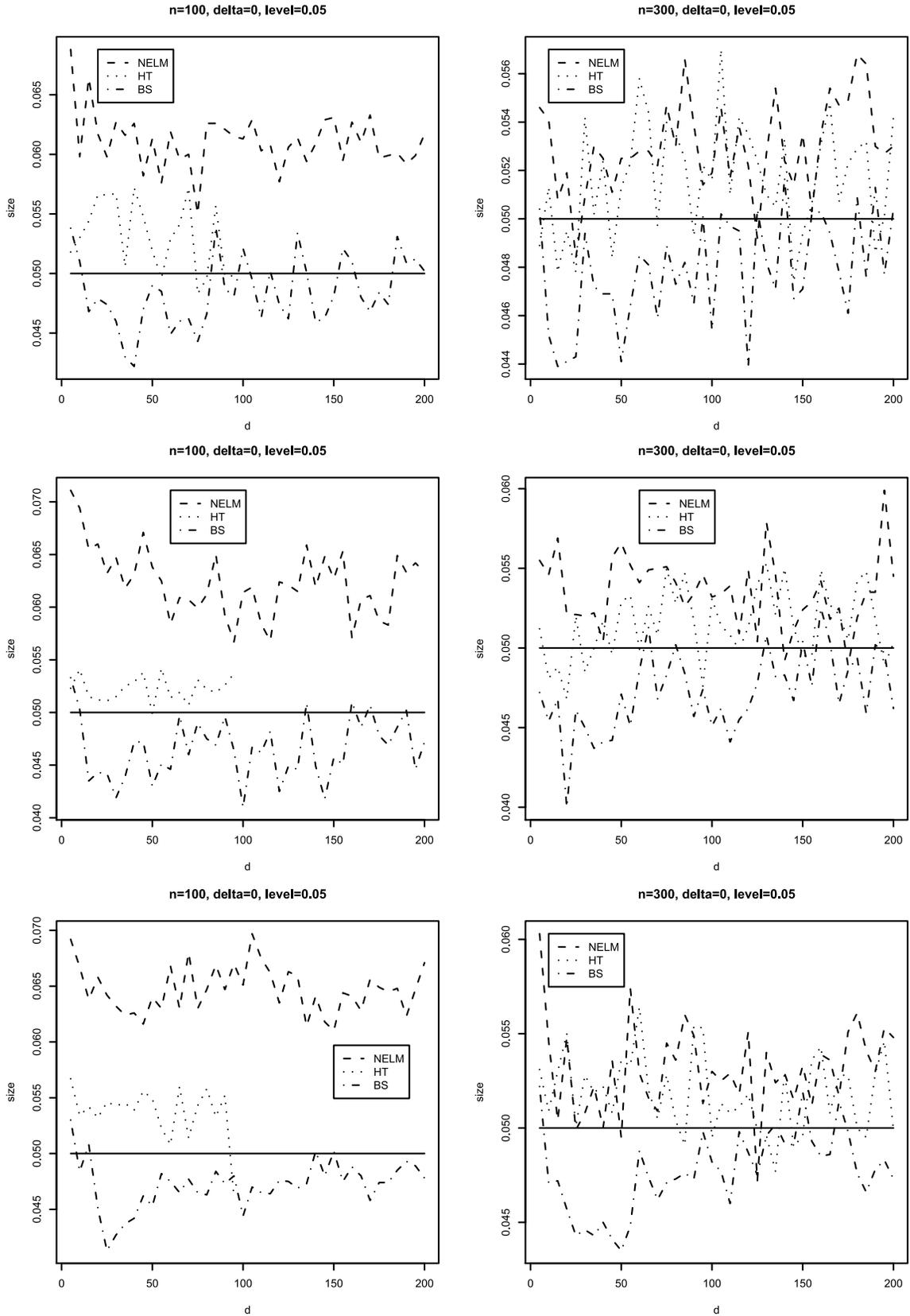}

\caption{Sizes of tests are plotted against $d=5,10,\ldots,200$ for
$\delta=0$ and level $0.05$. The upper, middle and lower panels
represent model~1 with $W_i\sim N(0,1)$, model 1 with $W_i\in t_6$ and
model 2, respectively. Solid line is the nominal level.}\label{fig1}
\end{figure*}

In Figure~\ref{fig1} we plot the empirical sizes (i.e.,
$\delta=0$) of these tests against $d=5,10,\ldots,200$ at a nominal
level $0.05$. Note that the Hotelling's $T^2$-test only works for
$d<n$. As we see, the size of the proposed empirical likelihood method is
slightly larger than the nominal level and less accurate than the other
two tests when $n=100$,
but it becomes close to the nominal level and comparable to the other
two tests
when $n=300$.

\begin{figure*}

\includegraphics{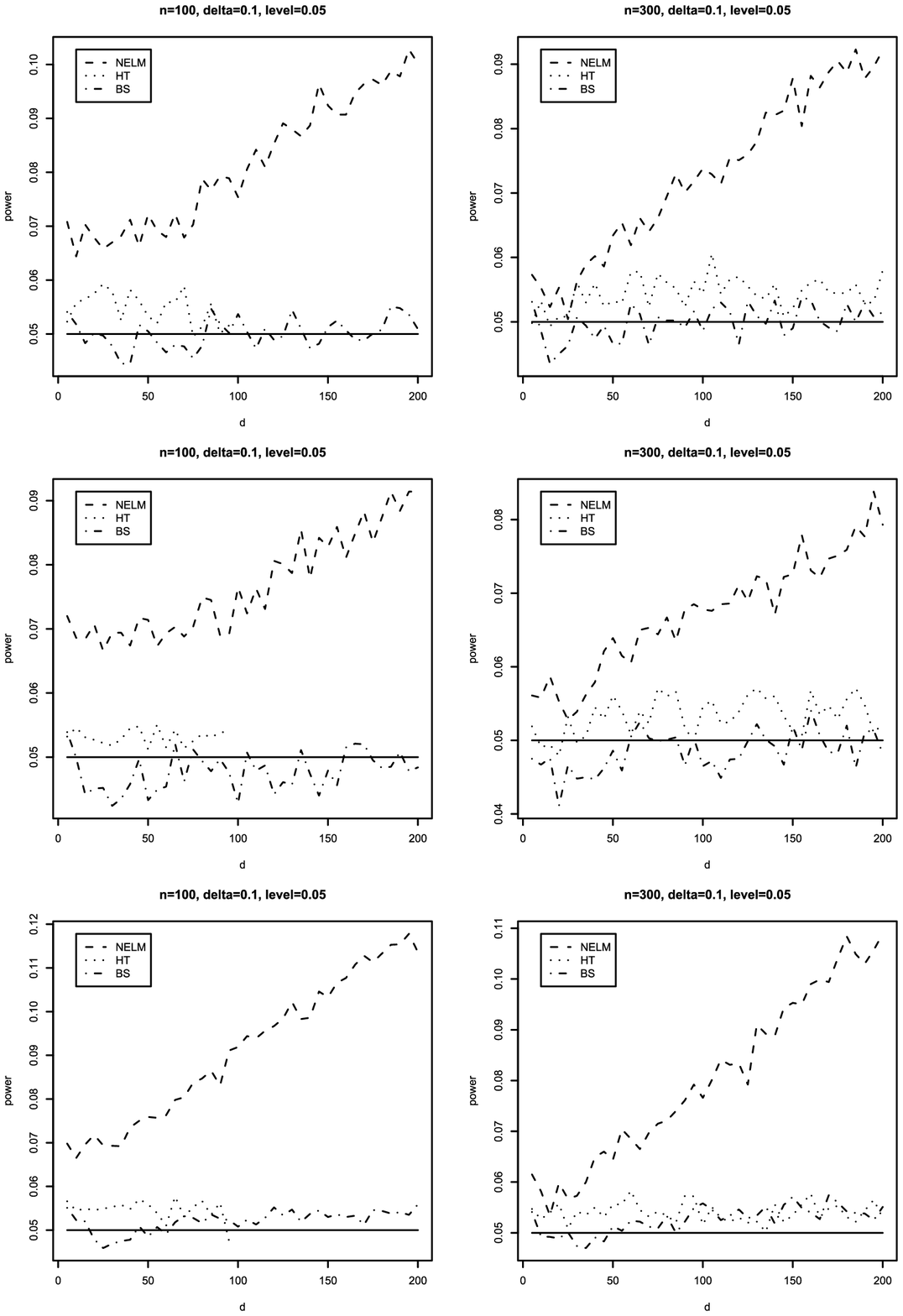}

\caption{Powers of tests are plotted against $d=5,10,\ldots,200$ for
$\delta=0.1$ and level $0.05$. The upper, middle and lower panels
represent model 1 with $W_i\sim N(0,1)$, model 1 with $W_i\in t_6$ and
model 2, respectively. Solid line is the nominal level.}\label{fig2}
\end{figure*}

\begin{figure*}

\includegraphics{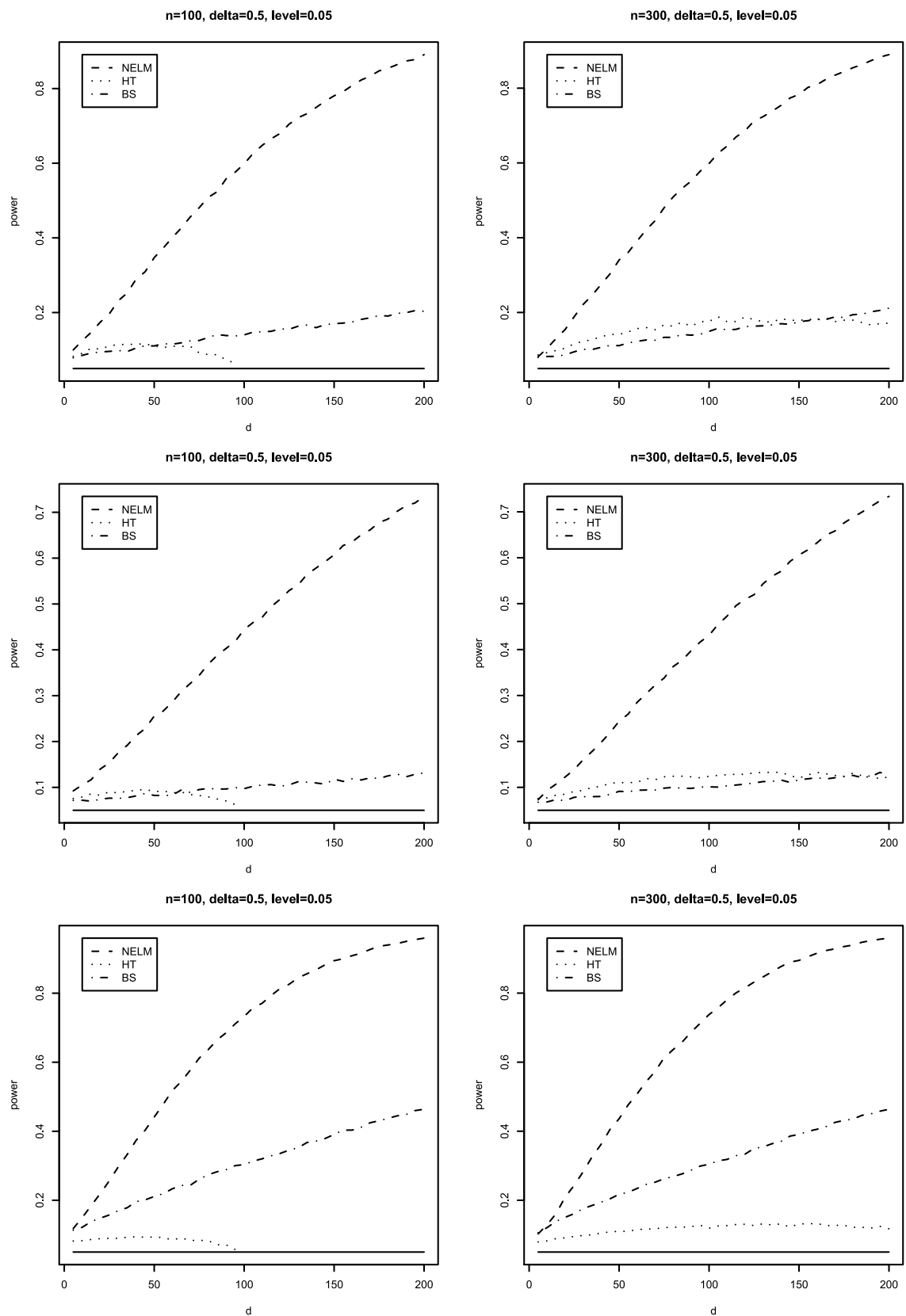}

\caption{Powers of tests are plotted against $d=5,10,\ldots,200$ for
$\delta=0.5$ and level $0.05$. The upper, middle and lower panels
represent model 1 with $W_i\sim N(0,1)$, model 1 with $W_i\in t_6$ and
model 2, respectively. Solid line is the nominal level.}\label{fig3}
\end{figure*}

In Figures~\ref{fig2} and \ref{fig3} the powers for $\delta=0.1$ and $0.5$ are plotted
against $d=5,10,\ldots,200$ at level $0.05$. These figures clearly
show that the proposed empirical likelihood method is much more
powerful than others especially when $d$ becomes relatively large.

In conclusion, the proposed empirical likelihood method has a stable
size with respect to the dimension and a large power, and performs well
for all considered~$d$.

\section{Proofs}\label{proof}

In the proofs we use $\|\cdot\|$ to denote the $L_2$ norm of a vector
or matrix.
Without loss of generality, we assume $\mu_0=0$. Write $u_i=u_i(0)$
and $v_i=v_i(0)$ for $1\le i \le m$.
Then it is easily verified that
\begin{eqnarray*}
E(u_1)&=&E(v_1)=E(u_1v_1)=0,
\\
\operatorname{Var}(u_1)&=&\sum^d_{i,j=1}
\sigma^2_{i,j}=:\pi_{11}
\end{eqnarray*}
and
\[
\operatorname{Var}(v_1)=2\sum^d_{i,j=1}
\sigma_{i,j}=:\pi_{22}.
\]

\begin{lemma}\label{lem1}
\begin{eqnarray*}
\operatorname{tr}(\Sigma^4)&=&O \bigl(\bigl(\mathbf
{tr}(\Sigma^2)\bigr)^2 \bigr),\\
\pi_{11}&=&\sum^d_{j=1}\lambda_j^2
\end{eqnarray*}
and
\[
2d\lambda_1\le\pi_{22}\le2d\lambda_d.
\]
\end{lemma}

\begin{pf}
Since $\operatorname{tr}(\Sigma^j)=\sum
^d_{i=1}\lambda_i^j$ for any positive integer $j$, the first equality
follows immediately.
The second equality follows since $\pi_{11}=\operatorname{tr}(\Sigma^2)$.
The third inequalities on $\pi_{22}$
can be proved easily. The proof of the lemma is complete.
\end{pf}

\begin{lemma}\label{lem2} Assume conditions (\ref{momentu0}) and
(\ref{momentv0}) hold.
Then
%
\begin{eqnarray}
\label{clt}
\frac{1}{\sqrt{m}}\sum^m_{i=1}
\pmatrix{\displaystyle \frac{u_i}{\sqrt{\pi
_{11}}}
\vspace*{2pt}\cr
\displaystyle \frac{v_i}{\sqrt{\pi_{22}}}} &\td& N(0, I_2),
\\
\label{weaku}
\frac{\sum^m_{i=1}u_i^2}{m\pi_{11}}-1&\tp&0,
\\
\label{weakv}
\frac{\sum^m_{i=1}v_i^2}{m\pi_{22}}-1&\tp&0,
\\
\label{weakuv}
\frac{\sum^m_{i=1}u_iv_i}{m\sqrt{\pi_{11}\pi_{22}}}&\tp&0,
\end{eqnarray}
where $I_2$ is a $2\times2$ identity matrix.
Moreover, we have
%
\begin{eqnarray}
\label{max} \max_{1\le i \le m}\biggl|\frac{u_i}{\sqrt{\pi_{11}}}\biggr|&=&o_p
\bigl(m^{1/2}\bigr) \quad\mbox{and}\nonumber\\[-8pt]\\[-8pt]
\max_{1\le i\le m}\biggl|
\frac{v_i}{\sqrt{\pi_{22}}}\biggr|&=&o_p\bigl(m^{1/2}\bigr).\nonumber
\end{eqnarray}
\end{lemma}

\begin{pf}
Note that $u_1$ and $v_1$ are uncorrelated.
To show (\ref{clt}), we need to prove that for any constants $a$ and
$b$ with $a^2+b^2\ne0$,
\[
\frac{1}{\sqrt{m}}\sum^m_{i=1}\biggl(a
\frac{u_i}{\sqrt{\pi_{11}}}+ b\frac{v_i}{\sqrt{\pi_{22}}}\biggr)\td
N\bigl(0, a^2+b^2
\bigr).
\]
Therefore, we shall verify the Lindeberg condition, which is a
consequence of the Lyapunov condition as follows:
%
\begin{eqnarray}
\label{lya}
&&\frac{1}{m^{(2+\delta)/2}}\sum^m_{i=1}E\biggl|a
\frac{u_i}{\sqrt{\pi_{11}}}+ b\frac{v_i}{\sqrt{\pi_{22}}}\biggr|^{2+\delta
}\nonumber\\
&&\quad= \frac{1}{m^{\delta/2}}E\biggl|a
\frac{u_1}{\sqrt{\pi_{11}}}+ b\frac
{v_1}{\sqrt{\pi_{22}}}\biggr|^{2+\delta}
\\
&&\quad\le\frac{(2|a|)^{2+\delta}}{m^{\delta/2}}E\biggl|\frac{u_1}{\sqrt{\pi
_{11}}}\biggr|^{2+\delta}\nonumber\\
&&\qquad{}+
\frac{(2|b|)^{2+\delta}}{m^{\delta/2}}E\biggl|\frac{v_1}{\sqrt{\pi
_{22}}}\biggr|^{2+\delta}
\nonumber\\
&&\quad\to 0
\nonumber
\end{eqnarray}
from conditions (\ref{momentu0}) and (\ref{momentv0}).

To show (\ref{weaku}), we need to estimate
$E|\sum^m_{i=1}u_i^2-m\pi_{11}|^{(2+\delta)/2}$. We have from
\citet{vonEss65} that
%
\begin{eqnarray}
\label{est1} &&
E\Biggl|\sum^m_{i=1}u_i^2-m
\pi_{11}\Biggr|^{(2+\delta)/2}\nonumber\\
&&\quad\le2mE\bigl|u_1^2-E
\bigl(u_1^2\bigr)\bigr|^{(2+\delta)/2}\\
&&\quad=O\bigl(mE|u_1|^{2+\delta}
\bigr),\nonumber
\end{eqnarray}
if $0<\delta\le2$, and from \citet{DhaJog69} that
%
\begin{eqnarray}
\label{est2}
&&
E\Biggl|\sum^m_{i=1}u_i^2-m
\pi_{11}\Biggr|^{(2+\delta)/2}\nonumber\\
&&\quad\le Cm^{(2+\delta)/4}E\bigl|u_1^2-E
\bigl(u_1^2\bigr)\bigr|^{(2+\delta)/2}\\
&&\quad=O\bigl(m^{(2+\delta
)/4}E|u_1|^{2+\delta}
\bigr),\nonumber
\end{eqnarray}
if $\delta>2$. Therefore, by (\ref{est1}), (\ref{est2}) and (\ref
{momentu0}) we have for any $\varepsilon>0$
\begin{eqnarray*}
&&P\biggl(\biggl|\frac{\sum^m_{i=1}u_i^2}{m\pi_{11}}-1\biggr|>\varepsilon\biggr)
\\
&&\quad\le \varepsilon^{-(2+\delta)/2}\frac{E|\sum^m_{i=1}u_i^2-m\pi
_{11}|^{(2+\delta)/2}}{(m\pi_{11})^{(2+\delta)/2}}
\\
&&\quad=O\biggl(m^{-(\delta+\min(\delta, 2))/4}E\biggl|\frac{u_1}{\sqrt{\pi
_{11}}}\biggr|^{2+\delta}\biggr)
\\
&&\quad=o(1),
\end{eqnarray*}
which implies (\ref{weaku}). Similarly, we can show (\ref{weakv}) and
(\ref{weakuv}). Equation (\ref{max}) follows from the Lyapunov
condition (\ref{lya}) by letting $a=1$ and $b=0$ or $a=0$ and $b=1$.
This completes the proof of the lemma.
\end{pf}

\begin{lemma}\label{lem3} For any $\delta>0$
\[
E|u_1|^{2+\delta}\le d^{\delta} \Biggl(\sum
^d_{i=1}E|X_{1,i}|^{2+\delta}
\Biggr)^2
\]
and
\[
E|v_1|^{2+\delta}\le2^{4+\delta}d^{1+\delta}\sum
^d_{i=1}E|X_{1,i}|^{2+\delta}.
\]
\end{lemma}

\begin{pf}
It follows from the Cauchy--Schwarz
inequality that
\[
|u_1|^2\le\|X_1\|^2\|X_{m+1}\|^2.
\]
Using the $C_r$ inequality that $E|\sum^d_{i=1}Z_i|^r\le d^{r-1}\*\sum
^d_{i=1}E|Z_i|^r$ for any
random variables $Z_1,\ldots, Z_d$ and positive constant $r>1$, we
conclude that
\begin{eqnarray*}
E|u_1|^{2+\delta}&\le& E \Biggl(\sum
^d_{i=1}X_{1,i}^2
\Biggr)^{\!\!(2+\delta)/2}\!E \Biggl(\sum^d_{i=1}X_{m+1, i}^2
\Biggr)^{\!\!(2+\delta
)/2}
\\
&=& \Biggl(E \Biggl(\sum^d_{i=1}X_{1,i}^2
\Biggr)^{(2+\delta)/2} \Biggr)^2
\\
&\le& \Biggl(d^{\delta/2}\sum^d_{i=1}E|X_{1,i}|^{2+\delta}
\Biggr)^2
\\
&=&d^{\delta} \Biggl(\sum^d_{i=1}E|X_{1,i}|^{2+\delta}
\Biggr)^2.
\end{eqnarray*}

Similarly, from the $C_r$ inequality we have
\begin{eqnarray*}
E|v_1|^{2+\delta}&\le&2^{4+\delta}E \Biggl(\sum
^d_{i=1}|X_{1,i}| \Biggr)^{2+\delta}\\
&\le&2^{4+\delta}d^{1+\delta}\sum^d_{i=1}E|X_{1,i}|^{2+\delta}.
\end{eqnarray*}
This completes the proof.
\end{pf}

\begin{pf*}{Proof of Theorem \ref{thm1}}
Set $u_i'=u_i/\sqrt{\pi _{11}}$, $v_i'=v_i/\sqrt{\pi_{22}}$ and
$Y_i'=(u_i', v_i')^T $ for $i=1,\ldots, m$. Then it is easy to see that
\[
\tilde l(0)=-2\log\tilde L(0)=2\sum_{i=1}^m
\log\bigl\{1+\rho^T Y_i'\bigr\},
\]
where $\rho=(\rho_1, \rho_2)^T $ solves
%
\begin{equation}
\label{Lag1} \frac1m\sum_{i=1}^m
\frac{Y_i'}{1+\rho^T Y_i'}=0.
\end{equation}

It follows from Lemma \ref{lem2} that
%
\begin{eqnarray}
\label{clt1}
\frac{1}{\sqrt{m}}\sum^m_{i=1}Y_i'
&\td& N(0, I_2),
\\
\label{weakmatrix}
\Biggl\|\frac1m\sum^m_{i=1}Y_i'
\bigl(Y_i'\bigr)^T -I_2\Biggr\|&\tp&0,
\\
\label{max1}
\max_{1\le i \le m}\bigl\|Y_i'\bigr\|&=&o_p
\bigl(m^{1/2}\bigr).
\end{eqnarray}

Similar to the proof of
(2.14) in \citet{Owe90}, we can show $\|\rho\|=O_p(m^{-1/2})$. Then it
follows from (\ref{max1}) that
\[
\max_{1\le i\le m}\biggl\|\frac{\rho^T Y_i'}{1+\rho^T Y_i'}\biggr\|=o_p(1).
\]
Therefore, we have from (\ref{Lag1}) that
\begin{eqnarray*}
0&=&\frac1m\sum_{i=1}^m\frac{\rho^T Y_i'}{1+\rho^T Y_i'}
\\
&=&\frac1m\sum_{i=1}^m\rho^T
Y_i'\biggl(1-\rho^T Y_i'+
\frac{(\rho^T Y_i')^2}{1+\rho^T Y_i'}\biggr)
\\
&=&\frac1m\sum_{i=1}^m\rho^T
Y_i'-\frac1m\sum^m_{i=1}
\bigl(\rho^T Y_i'\bigr)^2+
\frac{1}m\sum^m_{i=1}
\frac{(\rho^T Y_i')^3}{1+\rho
^T Y_i'}
\\
&=&\frac1m\sum_{i=1}^m\rho^T
Y_i'-\frac{(1+o_p(1))}m\sum
^m_{i=1} \bigl(\rho^T
Y_i'\bigr)^2,
\end{eqnarray*}
which implies
%
\begin{equation}
\label{eq1} \frac1m\sum_{i=1}^m
\rho^T Y_i'=\frac{(1+o_p(1))}m\sum
^m_{i=1} \bigl(\rho^T
Y_i'\bigr)^2.
\end{equation}
By using (\ref{Lag1}) and (\ref{weakmatrix}) we obtain
\begin{eqnarray*}
0&=&\frac1m\sum_{i=1}^m\frac{Y_i'}{1+\rho^T Y_i'}
\\
&=&\frac1m\sum_{i=1}^mY_i'
\biggl(1- \bigl(Y_i'\bigr)^T\rho+
\frac{(\rho^T Y_i')^2}{1+\rho^T Y_i'}\biggr)
\\
&=&\frac1m\sum_{i=1}^mY_i'-
\frac1m\sum^m_{i=1} Y_i'
\bigl(Y_i'\bigr)^T \rho+\frac{1}m
\sum^n_{i=1}\frac{Y_i'(\rho^T
Y_i')^2}{1+\rho^T Y_i'}
\\
&=&\frac1m\sum_{i=1}^mY_i'-
\frac1m\sum^m_{i=1} Y_i'
\bigl(Y_i'\bigr)^T \rho\\
&&{}+O_p
\Biggl(\max_{1\le i\le m}\biggl\|\frac{Y_i'}{1+\rho
^T Y_i'}\biggr\|\frac1{m}\sum
^n_{i=1}\bigl(\rho^T
Y_i'\bigr)^2 \Biggr)
\\
&=&\frac1m\sum_{i=1}^mY_i'-
\frac1m\sum^m_{i=1} Y_i'
\bigl(Y_i'\bigr)^T \rho\\
&&{}+o_p
\Biggl(m^{1/2}\rho^T \Biggl(\frac1{m}\sum
^n_{i=1}Y_i'
\bigl(Y_i'\bigr)^T \Biggr)\rho\Biggr)
\\
&=&\frac1m\sum_{i=1}^mY_i'-
\frac1m\sum^m_{i=1} Y_i'
\bigl(Y_i'\bigr)^T \rho+o_p
\bigl(m^{-1/2}\bigr),
\end{eqnarray*}
which implies
%
\begin{equation}
\label{eq2} \rho= \Biggl(\frac1m\sum^m_{i=1}
Y_i'\bigl(Y_i'
\bigr)^T \Biggr)^{-1}\frac1m\sum
_{i=1}^mY_i'+o_p
\bigl(m^{-1/2}\bigr).\hspace*{-28pt}
\end{equation}

Finally, by using Taylor's expansion, (\ref{eq1}), (\ref{eq2}),
(\ref{clt1}) and (\ref{weakmatrix}), we obtain
\begin{eqnarray*}
\tilde l(0)&=&2\sum^m_{i=1}
\rho^T Y_i'-\bigl(1+o_p(1)
\bigr)\sum^m_{i=1}\bigl(\rho^T
Y_i'\bigr)^2
\\
&=&\bigl(1+o_p(1)\bigr)\rho^T \Biggl(\sum
^m_{i=1}Y_i'
\bigl(Y_i'\bigr)^T \Biggr)\rho
\\
&=&\bigl(1+o_p(1)\bigr) \Biggl(\frac1{\sqrt{m}}\sum
_{i=1}^mY_i'
\Biggr)^T\\
&&\hspace*{0pt}{}\cdot \Biggl(\frac1m\sum^m_{i=1}Y_i'
\bigl(Y_i'\bigr)^T \Biggr)^{-1}
\frac1{\sqrt{m}}\sum_{i=1}^mY_i'\\
&&{}+o_p(1)
\\
&\stackrel{d}\to& \chi^2_2 \quad\mbox{as } n\to\infty.
\end{eqnarray*}
This completes the proof of Theorem \ref{thm1}.
\end{pf*}

\begin{pf*}{Proof of Corollary \ref{cor1}}
Equations (\ref{momentu0}) and (\ref{momentv0}) follow from conditions
(A1)--(A3) by using Lemmas \ref{lem1} and \ref{lem3}.
\end{pf*}

\begin{pf*}{Proof of Theorem \ref{thm2}}
It suffices to verify conditions (\ref{momentu0}) and (\ref{momentv0})
with $\delta=2$ in Theorem \ref{thm1}. As before, assume $\mu_0=0$.
Write $\Gamma=(\gamma_{i,j})_{1\le i\le d, 1\le j\le k}$.\vspace*{1pt} Then
$\operatorname{Var}(X_1)=\Sigma=\Gamma\Gamma^T$.
Denote $\done^T\Gamma =(a_1,\ldots,
a_k)$ and $\Sigma'=\Gamma^T\Gamma=(\sigma '_{j,l})_{1\le j, l\le k}$.
Then
$
v_1=v_1(0)=\sum_{j=1}^ka_j(Z_{1,j}+Z_{1+m,j})
$
and
\[
u_1=u_1(0)=\sum_{j=1}^k\sum_{l=1}^k\sigma_{j,l}'Z_{1,j}Z_{1+m,l}.
\]

Set $\delta_{j_1,j_2,j_3,j_4}=E(Z_{1,j_1}Z_{1,
j_2}Z_{1,j_3}Z_{1,j_4})$. Then
\[
\delta_{j_1,j_2,j_3,j_4}=3+\Delta,
\]
if $j_1=j_2=j_3=j_4$, $1$ if $j_1, j_2, j_3$ and $j_4$ form two
different pairs of integers, and zero otherwise. It follows from Lemma
\ref{lem1} that
\begin{eqnarray*}
Eu_1^4 &=&\sum_{j_1, j_2, j_3, j_4=1}^k
\sum_{l_1, l_2, l_3,
l_4=1}^k\sigma_{j_1,l_1}'
\sigma_{j_2,l_2}'\sigma_{j_3, l_3}'\sigma
_{j_4, l_4}' \\
&&\hspace*{94pt}{}\cdot\delta_{j_1,j_2,j_3,j_4}\delta_{l_1,l_2,l_3,l_4}
\\
&=&O \biggl(\biggl|\sum_{j_1\neq j_2}\sum
_{l_1\neq l_2}\sigma_{j_1,l_1}'
\sigma_{j_1,l_2}'\sigma_{j_2,l_1}'\sigma
_{j_2,l_2}'\biggr| \biggr)\\
&&{}+O \Biggl(\sum
_{j_1\neq j_2}\sum_{l=1}^k\sigma
_{j_1,l}'^2\sigma_{j_2,l}'^2
\Biggr)
\\
&&{}+O \Biggl(\sum_{j=1}^k\sum
_{l_1\neq l_2}\sigma_{j,l_1}'^2\sigma
_{j,l_2}'^2 \Biggr)+O \Biggl(\sum
_{j=1}^k\sum_{l=1}^k
\sigma_{j,l}'^4 \Biggr)
\\
&=&O \Biggl(\Biggl|\sum_{j_1=1}^k\sum
_{j_2=1}^k\sum_{l_1=1}^k
\sum_{l_2=1}^k\sigma_{j_1,l_1}'
\sigma_{j_1,l_2}'\sigma_{j_2,l_1}'\sigma
_{j_2,l_2}'\Biggr| \Biggr)
\\
&&{}+O \Biggl(\sum_{j_1=1}^k\sum
_{j_2=1}^k\sum_{l=1}^k
\sigma_{j_1,l}'^2\sigma_{j_2,l}'^2
\Biggr)\\
&&{}+O \Biggl(\sum_{j=1}^k\sum
_{l=1}^k\sigma_{j,l}'^4
\Biggr)
\\
&=&O \bigl(\operatorname{tr}\bigl(\Sigma'^4\bigr) \bigr)+O
\Biggl(\Biggl(\sum_{j=1}^k\sum
_{l=1}^k\sigma_{j,l}'^2
\Biggr)^2\Biggr)
\\
&=&O \bigl(\operatorname{tr}\bigl(\Sigma'^4\bigr) \bigr)+O
\bigl(\bigl(\operatorname{tr}\bigl(\Sigma'^2\bigr)
\bigr)^2 \bigr)
\\
&=&O \bigl(\operatorname{tr}\bigl(\Sigma^4\bigr) \bigr)+O \bigl(\bigl(
\operatorname{tr}\bigl(\Sigma^2\bigr)\bigr)^2 \bigr)
\\
&=&o \bigl(m\bigl(\operatorname{tr}\bigl(\Sigma^2\bigr)
\bigr)^2 \bigr),
\end{eqnarray*}
that is, (\ref{momentu0}) holds with $\delta=2$.

Similarly, we have
\begin{eqnarray*}
Ev_1^4&\le&2^4E \Biggl(\sum
_{j=1}^ka_jZ_{1,j}
\Biggr)^4
\\
&=&O \Biggl( \sum_{j_1, j_2=1}^ka_{j_1}^2a_{j_2}^2
\Biggr)+O \Biggl(\sum_{j=1}^ka_j^4
\Biggr)
\\
&=&O \Biggl( \Biggl(\sum_{j=1}^ka_j^2
\Biggr)^2 \Biggr)
\\
&=&O \bigl( \bigl(\done^T\Gamma\Gamma^T\done
\bigr)^2 \bigr)
\\
&=&O \Biggl( \Biggl(\sum^d_{i=1}\sum
^d_{j=1}\sigma_{i,j}
\Biggr)^2 \Biggr),
\end{eqnarray*}
which yields (\ref{momentv0}) with $\delta=2$. The proof is complete.
\end{pf*}

\begin{pf*}{Proof of Theorem \ref{thm3}}
We continue to use the notation in the proof of Theorem \ref{thm1}.
Define
\begin{eqnarray*}
\rho_{n1}&=&\frac{(\mu_0-\mu)^T(\mu_0-\mu)}{\sqrt{\pi_{11}}},\\
\rho_{n2}&=&
\frac{{\mathbf1}_d^T(2\mu-2\mu_0)}{\sqrt{\pi_{22}}}.
\end{eqnarray*}
Then $\tau=m\rho_{n1}^2+m\rho_{n2}^2$.

Notice that the true value for the mean of $X_1$ is $\mu$
under the alternative hypothesis.
Since for $1\le i\le m$
%
\begin{eqnarray}
\label{uuu} u_i(\mu_0)&=&u_i(\mu)
+(\mu-
\mu_0)^T(\mu-\mu_0)\nonumber\\[-8pt]\\[-8pt]
&&{}+(\mu-\mu
_0)^T(X_i+X_{i+m}-2\mu)\nonumber
\end{eqnarray}
and
\[
v_i(\mu_0)=v_i(\mu)+{
\mathbf1}_d^T(2\mu-2\mu_0),
\]
we have
\[
Y_i'=\pmatrix{u_i(\mu)/\sqrt{
\pi_{11}}
\cr
v_i(\mu)/\sqrt{\pi_{22}}}+\pmatrix{
\rho_{n1}
\cr
\rho_{n2}}+ \pmatrix{s_i(\mu)
\cr
0},
\]
where $s_i(\mu)=(\mu-\mu_0)^T(X_i+X_{i+m}-2\mu)/\sqrt{\pi_{11}}$
and $Y_i'={u_i(\mu_0)/\sqrt{\pi_{11}}\choose v_i(\mu
_0)/\sqrt{\pi_{22}}}$ as defined in the proof of Theorem~\ref{thm1}.

First we consider the case of $\tau=o(m)$.
Since $\tau=o(m)$ implies that $\rho_{n1}=o(1)$ and $\rho
_{n2}=o(1)$, it follows from Lemma \ref{lem1} that
%
\begin{eqnarray}
\label{add0} E\bigl(s_1^2(\mu)\bigr)&=&O \biggl(
\frac{1}{\pi_{11}}(\mu-\mu_0)^T\Sigma(\mu-
\mu_0) \biggr)
\nonumber
\\
&=&O \biggl(\frac{\lambda_d}{\pi_{11}}(\mu-\mu_0)^T (\mu-\mu
_0) \biggr)
\\
&=&O\biggl(\frac{\lambda_d}{\sqrt{\pi_{11}}}\rho_{n1}\biggr)\to0,
\nonumber
\end{eqnarray}
which implies
\[
\frac{\sum^m_{i=1}s_i^2(\mu)}{m}\tp0
\]
and
\[
\frac{\max_{1\le i\le m}|s_i(\mu)|}{\sqrt{m}}\le\sqrt{\frac{\sum
^m_{i=1}s_i^2(\mu)}{m}}\tp0
\]
as $m\to\infty$. Hence, we conclude that
%
\begin{eqnarray}
\label{clt2} V_n&:=&\pmatrix{V_{n1}
\cr
V_{n2}}=
\frac{1}{\sqrt{m}}\sum^m_{i=1}
\left\{Y_i'- \pmatrix{\rho_{n1}
\cr
\rho_{n2}} \right\}\nonumber\\[-8pt]\\[-8pt]
&\td& N(0, I_2),\nonumber
\end{eqnarray}
and both (\ref{weakmatrix}) and (\ref{max1}) hold when $\tau=o(m)$.
Following the
proof of Theorem \ref{thm1}, we can show that
%
\begin{eqnarray}
\label{add1} \tilde l(\mu_0)&=&\bigl(1+o_p(1)\bigr)
\Biggl(\frac1{\sqrt{m}}\sum_{i=1}^mY_i'
\Biggr)^T\nonumber\\
&&\hspace*{0pt}{}\cdot \Biggl(\frac1m\sum^m_{i=1}Y_i'
\bigl(Y_i'\bigr)^T \Biggr)^{-1}
\frac1{\sqrt{m}}\sum_{i=1}^mY_i'+o_p(1)
\nonumber\\[-8pt]\\[-8pt]
&=&(V_{n1}+\sqrt{m}\rho_{n1})^2
\bigl(1+o_p(1)\bigr)\nonumber\\
&&{}+(V_{n2}+\sqrt{m}\rho
_{n2})^2\bigl(1+o_p(1)\bigr)
+o_p(1),
\nonumber
\end{eqnarray}
when $\tau=o(m)$.

If the limit of $\tau$, say, $\tau_0$, is finite, then it follows
from (\ref{clt2}) and (\ref{add1}) that
$\tilde l(\mu_0)$ converges in distribution to a noncentral chi-square
distribution with two degrees of freedom and noncentrality parameter
$\tau_0$
and, consequently, (\ref{power}) holds. If $\tau$ goes to infinity,
the limit of the right-hand side of (\ref{power}) is 1.
By (\ref{add1}), together with the elementary inequality $(a+b)^2\ge
\frac{a^2}{2}-b^2$, we have that
%
\begin{eqnarray}
\label{add1a} \tilde l(\mu_0)&\ge&\biggl(\frac{m\rho_{n1}^2}2-V_{n1}^2
\biggr) \bigl(1+o_p(1)\bigr)\nonumber\\
&&{}+\biggl(\frac
{m\rho_{n2}^2}2-V_{n2}^2
\biggr) \bigl(1+o_p(1)\bigr)+o_p(1)
\nonumber
\\
&=&\frac{\tau}{2}\bigl(1+o_p(1)\bigr)\\
&&{}-\bigl(V_{n1}^2+V_{n2}^2
\bigr) \bigl(1+o_p(1)\bigr)+o_p(1)
\nonumber\\
&\stackrel{p} {\to}&\infty,
\nonumber
\end{eqnarray}
which implies that the limit of the left-hand side of (\ref{power}) is
also 1. Thus, (\ref{power}) also holds when $\tau=o(m)$.

Next we consider the case of $\lim\inf\rho_{n2}^2>0$. Since $\sum
_{i=1}^mp_iY_i(\mu_0)=0$ implies that $\sum_{i=1}^mp_iv_i(\mu_0)=0$,
we have
%
\begin{eqnarray}
\label{add2} \tilde L(\mu_0)&\le&\sup\Biggl\{\prod
_{i=1}^m(mp_i)\dvtx p_1\ge0,\ldots,p_m\ge0,\nonumber\\
&&\hspace*{34.6pt} \sum_{i=1}^mp_i=1,
\sum_{i=1}^mp_iv_i(
\mu_0)=0\Biggr\}
\nonumber\\[-8pt]\\[-8pt]
&=&\sup\Biggl\{\prod_{i=1}^m(mp_i)\dvtx p_1
\ge0,\ldots,p_m\ge0,\nonumber\\
&&\hspace*{32.2pt}\sum_{i=1}^mp_i=1,
\sum_{i=1}^mp_i
\frac{v_i(\mu_0)}{\sqrt{\pi_{22}}}=0\Biggr\}.\nonumber
\end{eqnarray}
Define
\begin{eqnarray*}
L^*(\theta)&=&\sup\Biggl\{\prod_{i=1}^m(mp_i)\dvtx
p_1\ge0,\ldots,p_m\ge0,\\
&&\hspace*{23pt}\sum
_{i=1}^mp_i=1,\sum
_{i=1}^mp_i\biggl(\frac{v_i(\mu_0)}{\sqrt{\pi
_{22}}}-
\rho_{n2}\biggr)=\theta\Biggr\}.
\end{eqnarray*}
Put $\theta^*=\frac{1}m\sum_{i=1}^m(\frac{v_i(\mu_0)}{\sqrt{\pi
_{22}}}-\rho_{n2})$. Then
%
\begin{equation}
\label{add30} -2\log L^*\bigl(\theta^*\bigr)=0.
\end{equation}
Since $E\{v_i(\mu_0)/\sqrt{\pi_{22}}-\rho_{n2}\}=E\{v_i(\mu)/\sqrt{\pi
_{22}}\}=0$
and $E\{v_i(\mu_0)/\sqrt{\pi_{22}}-\rho_{n2}\}^2=1$ under $H_a\dvtx  \mu
\neq\mu_0$, we have by using Chebyshev's inequality that
%
\begin{equation}
\label{add3}P\bigl(\bigl|\theta^*\bigr|>m^{-2/5}\bigr)\to0.
\end{equation}
Using $E\{v_i(\mu_0)/\sqrt{\pi_{22}}-\rho_{n2}\}^2=1$, similar to
the proof of (\ref{add1a}), we can show that
\[
-2\log L^*\bigl(\theta^*_1\bigr)\stackrel{p}
{\to}\infty\quad\mbox{and}\quad
{-}2\log L^*\bigl(\theta^*_2\bigr)\stackrel{p} {\to}\infty,
\]
where $\theta^*_1=m^{-1/4}$ and $\theta^*_2=-m^{-1/4}$, which satisfy
$m(\theta^*_1)^2=o(m)$ and $m(\theta^*_2)^2=o(m)$. To help understand
this better, we first notice that $\tilde L(\mu_0)$ can be rewritten
as follows:
\begin{eqnarray*}
\tilde L(\mu_0)&=&\sup\Biggl\{\prod_{i=1}^m(mp_i)\dvtx p_1
\ge0,\ldots,p_m\ge0,\\
&&\hspace*{53.6pt} \sum_{i=1}^mp_i=1,
\sum_{i=1}^mp_iY_i'=0
\Biggr\}
\\
&=&\sup\Biggl\{\prod_{i=1}^m(mp_i)\dvtx p_1
\ge0,\ldots,p_m\ge0, \\
&&\hspace*{21.8pt}\sum_{i=1}^mp_i=1,\\
&&\hspace*{21.8pt}
\sum_{i=1}^mp_i
\left(Y_i'- \pmatrix{\rho_{n1}
\cr
\rho
_{n2}}\right) = -\pmatrix{\rho_{n1}
\cr
\rho_{n2}}
\Biggr\}.
\end{eqnarray*}
In equation (\ref{add1a}), it is the quantity
$m\|{-{\rho_{n1}\choose\rho_{n2}}}\|^2=\tau$ that
determines whether $\tilde l(\mu_0)$ diverges. As a one-dimensional
analogue of
$\tilde L(\mu_0)$, for any sequence $\theta_n$, if $\theta_n=o(1)$,
$-2\log L^*(\theta_n)$ can be expanded as in (\ref{add1}) via replacing
$Y_i'$ by $v_i(\mu_0)/\sqrt{\pi_{22}}-\rho_{n2}$, and replacing
$-{\rho_{n1}\choose\rho_{n2}}$ by $\theta_n$. And
if, further,
$m\theta_n^2\to\infty$, $-2\log L^*(\theta_n)$ goes to infinity in
probability, just like (\ref{add1a}). Obviously, with the choices of
$\theta_n=\pm m^{-1/4}$, conditions $\theta_n=o(1)$ and $m\theta
_n^2\to\infty$ are satisfied.

It follows from \citet{HalLaS90} that the set $\{\theta\dvtx
-2\log L^*(\theta)\le c\}=:I_c$ is convex for any $c$.
Take $c=c_n=\min\{-2\log L^*(\theta_1^*), -2\log L^*(\theta_2^*)\}
/2$. If $-\rho_{n2}$ belongs to the above convex set, then it follows
from (\ref{add30}) that $-a\rho_{n2}+(1-a)\theta^*$ belongs to that
convex set for any $a\in[0, 1]$ or, equivalently, any number between
$-\rho_{n2}$ and $\theta^*$ belongs to $I_{c_n}$. Recall that we
assume $\liminf\rho_{n2}^2>0$. Assume $n$ is large such that
$m^{-1/4}<|\rho_{n2}|$.
Under the condition $|\theta^*|\le m^{-2/5}$, if $\rho_{n2}>0$,
then $-\rho_{n2}<\theta^*_2=-m^{-1/4}<\theta^*$, and if $\rho
_{n2}<0$, then $\theta^*<\theta^*_1=m^{-1/4}<-\rho_{n2}$. Therefore,
if $|\theta^*|\le m^{-2/5}$ and $-\rho_{n2}\in I_{c_n}$, at least one
of $\theta^*_1$ and $\theta^*_2$ belongs to $I_{c_n}$. Precisely, we have,
as $n$ goes to infinity,
\begin{eqnarray*}
&&P\bigl(\bigl|\theta^*\bigr|\le m^{-2/5}, -\rho_{n2}\in
I_{c_n}\bigr)\\
&&\quad\le P\bigl(\theta_1^*\in I_{c_n}
\mbox{ or } \theta_2^*\in I_{c_n}\bigr)
\\
&&\quad=P\bigl(\min\bigl\{-2\log L^*\bigl(\theta_1^*\bigr), -2\log L^*
\bigl(\theta_2^*\bigr)\bigr\}\le c_n\bigr)\\
&&\quad=P\bigl(\min
\bigl\{-2\log L^*\bigl(\theta_1^*\bigr), -2\log L^*\bigl(
\theta_2^*\bigr)\bigr\}=0\bigr)\\
&&\quad\to0,
\end{eqnarray*}
which, together with (\ref{add3}), implies
\begin{eqnarray*}
&&
P\bigl( -2\log L^*(-\rho_{n2})>c_n\bigr)\\
&&\quad= P(-
\rho_{n2}\notin I_{c_n})\\
&&\quad\ge1- P\bigl(\bigl|\theta^*\bigr|\le
m^{-2/5}, -\rho_{n2}\in I_{c_n}\bigr)\\
&&\qquad{}-P\bigl(\bigl|
\theta^*\bigr|> m^{-2/5}\bigr)\\
&&\quad\to1
\end{eqnarray*}
and, therefore,
%
\begin{equation}
\label{add33} -2\log L^*(-\rho_{n2})\stackrel{p} {\to}\infty
\end{equation}
since $c_n\stackrel{p}{\to}\infty$.
Hence, combining with (\ref{add2}), we have
\begin{eqnarray*}
&&
P\bigl(-2\log\tilde L(\mu_0)>\xi_{1-\alpha}\bigr)\\
&&\quad\ge P\bigl(-2
\log L^*(-\rho_{n2})>\xi_{1-\alpha}\bigr)\\
&&\quad\to1,
\end{eqnarray*}
when $\lim\inf\rho_{n2}^2>0$.

Next we consider the case of $\lim\inf\rho_{n1}>0$. Define $\pi
_{33}=E\{(\mu-\mu_0)^T(X_1+X_{1+m}-2\mu)\}^2$ and $\rho_{n3}=\frac
{(\mu_0-\mu)^T(\mu_0-\mu)}{\sqrt{\pi_{11}+\pi_{33}}}$. As
before, we have
%
\begin{eqnarray}
\label{add4} \tilde L(\mu_0)&\le&\sup\Biggl\{\prod
_{i=1}^m(mp_i)\dvtx p_1\ge0,\ldots,p_m\ge0,\nonumber\\
&&\hspace*{33.5pt} \sum_{i=1}^mp_i=1,
\sum_{i=1}^mp_iu_i(
\mu_0)=0\Biggr\}
\nonumber\\[-8pt]\\[-8pt]
&=&\sup\Biggl\{\prod_{i=1}^m(mp_i)\dvtx p_1
\ge0,\ldots,p_m\ge0,\nonumber\\
&&\hspace*{22pt} \sum_{i=1}^mp_i=1,
\sum_{i=1}^mp_i
\frac{u_i(\mu_0)}{\sqrt{\pi_{11}+\pi
_{33}}}=0\Biggr\}.
\nonumber
\end{eqnarray}
Define
\begin{eqnarray*}
L^{**}(\theta)&=&\sup\Biggl\{\prod_{i=1}^m(mp_i)\dvtx
p_1\ge0,\ldots,p_m\ge0,\\
&&\hspace*{22pt} \sum
_{i=1}^mp_i=1,\\
&&\hspace*{22pt} \sum
_{i=1}^mp_i\biggl(\frac{u_i(\mu_0)}{\sqrt{\pi
_{11}+\pi_{33}}}-
\rho_{n3}\biggr)=\theta\Biggr\}.
\end{eqnarray*}

Since $u_1(\mu)$ and $(\mu-\mu_0)^T(X_1+X_{1+m}-2\mu)$ are
two uncorrelated variables with zero means, we have $\operatorname{Var}(u_1(\mu)+(\mu
-\mu_0)^T(X_1+X_{1+m}-2\mu))=\pi_{11}+\pi_{33}$.
As we have shown in the proof of Theorem \ref{thm2},
$E|u_1(\mu)|^4=o(m\pi_{11}^2)$. Following the same lines for
estimating $E(v_1^4)$ in the end of the proof of Theorem \ref{thm2}, we
have
\[
E\bigl\{(\mu-\mu_0)^T(X_1+X_{1+m}-2
\mu)\bigr\}^4=O\bigl(\pi_{33}^2\bigr).
\]
Then it follows that
\begin{eqnarray*}
&&
E\bigl\{u_1(\mu)+(\mu-\mu_0)^T(X_1+X_{1+m}-2
\mu)\bigr\}^4 \\
&&\quad\le 8\bigl(E\bigl|u_1(\mu)\bigr|^4\\
&&\qquad\hspace*{6.5pt}{}+E
\bigl\{(\mu-\mu_0)^T(X_1+X_{1+m}-2
\mu)\bigr\}^4\bigr)
\\
&&\quad=o\bigl(m(\pi_{11}+\pi_{33})^2\bigr).
\end{eqnarray*}
From (\ref{uuu}),
\begin{eqnarray*}
&&
\frac{u_i(\mu_0)}{\sqrt{\pi_{11}+\pi_{33}}}-\rho_{n3}\\
&&\quad=\frac
{u_i(\mu)+(\mu-\mu_0)^T(X_i+X_{i+m}-2\mu)}{\sqrt{\pi_{11}+\pi_{33}}}
\end{eqnarray*}
and, thus, we have
\begin{eqnarray*}
&&
E\biggl(\frac{u_i(\mu_0)}{\sqrt{\pi_{11}+\pi_{33}}}-\rho_{n3}\biggr)^4\\
&&\quad=
\frac
{E(u_i(\mu)+(\mu-\mu_0)^T(X_i+X_{i+m}-2\mu))^4}{(\pi_{11}+\pi
_{33})^2}\\
&&\quad=o(m).
\end{eqnarray*}
This ensures the validity of the Wilks theorem for $-2\log L^{**}(0)$,
that is,
$-2\log L^{**}(0)$ converges in distribution to a chi-square
distribution with one degree of freedom. Note that in Theorem \ref
{thm1}, two similar
conditions, (\ref{momentu0}) and (\ref{momentv0}), are imposed to
obtain the Wilks theorem for the log-empirical likelihood statistic for
two-dimensional mean vectors.
Similar to the proof of (\ref{add1a}), we can show that
\[
-2\log L^{**}\bigl(\theta^*_1\bigr)\stackrel{p} {\to}
\infty\quad\mbox{and}\quad {-}2\log L^{**}\bigl(\theta^*_2\bigr)
\stackrel{p} {\to}\infty,
\]
where\vspace*{1pt} $\theta^*_1=m^{-1/4}$ and $\theta^*_2=-m^{-1/4}$, which satisfy
$m(\theta^*_1)^2=o(m)$ and $m(\theta^*_2)^2=o(m)$.

Put $\theta^{**}=\frac{1}m\sum_{i=1}^m(\frac{u_i(\mu_0)}{\sqrt{\pi
_{11}+\pi_{33}}}-\rho_{n3})$. Then
%
\begin{equation}
\label{add50} -2\log L^{**}\bigl(\theta^{**}\bigr)=0.
\end{equation}
Since
\begin{eqnarray*}
&&E\bigl\{u_i(\mu_0)/\sqrt{\pi_{11}+
\pi_{33}}-\rho_{n3}\bigr\}\\
&&\quad=E\biggl\{\frac
{u_i(\mu)+(\mu-\mu_0)^T(X_i+X_{i+m}-2\mu)}{\sqrt{\pi_{11}+\pi
_{33}}}\biggr
\}\\
&&\quad=0
\end{eqnarray*}
and
\begin{eqnarray*}
&&E\biggl\{\frac{u_i(\mu_0)}{\sqrt{\pi_{11}+\pi_{33}}}-\rho_{n3}\biggr\}
^2\\
&&\quad=E\biggl\{
\frac{u_i(\mu)+(\mu-\mu_0)^T(X_i+X_{i+m}-2\mu)}{\sqrt{\pi
_{11}+\pi_{33}}}\biggr\}^2\\
&&\quad=1
\end{eqnarray*}
under $H_a\dvtx  \mu\neq\mu_0$,
we have from Chebyshev's inequality that
%
\begin{equation}
\label{add5}P\bigl(\bigl|\theta^{**}\bigr|>m^{-2/5}\bigr)\to0.
\end{equation}
By (\ref{add0}), we have
$
\pi_{33}/\pi_{11}=Es_1^2(\mu)=O(\rho_{n1})$, which implies that
there exists a constant $M>0$ such that
\begin{eqnarray*}
\rho_{n3}/m^{-1/4}&=&m^{1/4}\rho_{n1}
\frac{\sqrt{\pi_{11}}}{\sqrt{\pi_{11}+\pi_{33}}}\\
&\ge& m^{1/4}\rho_{n1}\{1+M
\rho_{n1}\}^{-1/2}\\
&\to&\infty
\end{eqnarray*}
since $\lim\inf\rho_{n1}>0$.

Using (\ref{add50}), (\ref{add5}) and the same arguments in proving
(\ref{add33}), we have
\[
-2\log L^{**}(-\rho_{n3})\stackrel{p} {\to}\infty.
\]
Hence, combining with (\ref{add4}), we have
\begin{eqnarray*}
&&P\bigl(-2\log\tilde L(\mu_0)>\xi_{1-\alpha}\bigr)\\
&&\quad\ge P\bigl(-2
\log L^{**}(-\rho_{n3})>\xi_{1-\alpha}\bigr)\\
&&\quad\to1,
\end{eqnarray*}
when\vspace*{1pt} $\lim\inf\rho_{n1}^2>0$.
Therefore, (\ref{power}) holds when $\lim\inf\rho_{n1}>0$.
This completes the proof of Theorem~\ref{thm3}.
\end{pf*}

\section*{Acknowledgments}

We thank the Editor, one Associate Editor and two reviewers for their
helpful comments. Peng's research was supported by NSF Grant
DMS-10-05336, Qi's research was supported by NSF Grant DMS-10-05345 and
Wang's research was supported by NSFC Grant No. 11271033, Foundation of
Beijing Education Bureau Grant No. KM201110028003.



\end{document}